\documentclass{article}
\makeatletter
\let\@fnsymbol\@arabic
\makeatother

\usepackage[letterpaper, margin=1.60in]{geometry}

\usepackage{graphicx}

\usepackage[colorlinks=true,
            urlcolor=blue,
            citecolor=blue,
            linkcolor=blue,
            bookmarks=false,
            bookmarksnumbered,
            ]{hyperref}

\newcommand{\NW}{\textsf{noWorkflow}}
\newcommand{\YW}{\textsf{YesWorkflow}}
\newcommand{\yw}{\textsf{YW}}

\newcommand{\YWT}{\textsf{YesWorkflow}}
\newcommand{\ywt}{\textsf{YW}}

\newcommand{\ywa}[1]{\texttt{#1}}
\newcommand{\ywm}[1]{\texttt{#1}}

\newcommand{\R}{\textsf{R}}
\newcommand{\MATLAB}{\textsf{MATLAB}}

\newcommand{\figref}[1]{Figure\,\ref{#1}}

\usepackage{etoolbox}
\patchcmd{\author}{0.45}{0.9}{}{}

\begin{document}

\title{\bf \YWT: A User-Oriented, Language-Independent Tool for Recovering Workflow Information from Scripts}

\author{
  Timothy McPhillips\footnote{Graduate School for Library and
    Information Science (GSLIS), University of Illinois at
    Urbana-Champaign (UIUC);
   $^2$Dept.\ of Computer Science, University of California, Davis;
   $^3$University of Calgary;
   $^4$University Corporation for Atmospheric Research (UCAR) and U.S.
   Global Change Research Program (USGCRP);
  $^5$Paris Dauphine University, LAMSADE;
  $^6$Department of Anthropology, Washington State University, Pullman, WA; 
  $^7$New York University;
  $^8$University of California, Santa Barbara;
  $^9$University of Massachusetts, Dartmouth;
  $^{10}$University of Newcastle, UK;
  $^{11}$Northern Arizona University;
  $^{12}$Oak Ridge National Laboratoy;
  $^{13}$University of Edinburgh, Scotland;
  $^{14}$National Center for Advanced Supercomputing
   Applications (NCSA), UIUC.} \and
  \and 
  Tianhong Song$^2$ \and
  Tyler Kolisnik$^3$ \and 
  Steve Aulenbach$^4$ \and 
  Khalid Belhajjame$^5$ \and
  Kyle Bocinsky$^6$ \and
  Yang Cao$^1$ \and 
  Fernando Chirigati$^7$ \and
  Saumen Dey$^2$ \and
  Juliana Freire$^7$ \and   
  Deborah Huntzinger$^{11}$ \and
  Christopher Jones$^8$  \and 
  David Koop$^9$ \and 
  Paolo Missier$^{10}$ \and 
  Mark Schildhauer$^8$ \and 
  Christopher Schwalm$^{11}$ \and
  Yaxing Wei$^{12}$ \and
   James Cheney$^{13}$ \and
  Mark Bieda$^3$ \and
  Bertram Lud\"ascher$^{1, 14}$
}

\date{\relax}

\maketitle

\begin{abstract}
  Scientific workflow management systems offer features for composing
  complex computational pipelines from modular building blocks, for
  executing the resulting automated workflows, and for recording the
  provenance of data products resulting from workflow runs.  Despite
  the advantages such features provide, many automated workflows
  continue to be implemented and executed outside of scientific
  workflow systems due to the convenience and familiarity of scripting
  languages (such as Perl, Python, \R, and \MATLAB), and to the high
  productivity many scientists experience when using these languages.
  \YW\ is a set of software tools that aim to provide such users of
  scripting languages with many of the benefits of scientific workflow
  systems.  \YW\ requires neither the use of a workflow engine nor the
  overhead of adapting code to run effectively in such a system.
  Instead, \YW\ enables scientists to annotate existing scripts with
  special comments that reveal the computational modules and dataflows
  otherwise implicit in these scripts.  \YW\ tools extract and analyze
  these comments, represent the scripts in terms of entities based on
  the typical scientific workflow model, and provide graphical
  renderings of this workflow-like view of the scripts. Future versions
of \YW\ also will allow the prospective provenance of the data products of 
  these scripts to be queried in ways similar to those available to users 
  of scientific workflow systems.
\end{abstract}

\section{Introduction}

Many scientists use scripts (written, e.g., in Python, \R, or \MATLAB)
or scientific workflow environments for data processing, analysis,
model simulation, result visualization, and other scientific computing
tasks. In addition to the widespread use in the natural sciences,
computational automation tools are also increasingly used in other
domains, e.g., for data mining workflows in the digital humanities
\cite{van2012if}, or to implement data curation workflows for natural
history collections \cite{Dou2012kurator}. One advantage of using
\emph{scientific workflow systems} (e.g., Galaxy \cite{Goecks2010},
Kepler \cite{ludascher2006scientific}, Taverna \cite{oinn2004taverna},
VisTrails \cite{bavoil2005vistrails}, RestFlow
\cite{mcphillips2010restflow,tsai2013autodrug}) 
is that they often include capabilities to track data as it is being
processed. By capturing and subsequently sharing such
\emph{provenance} information, scientists can provide a detailed
account of how their results were derived from the given inputs via
intermediate results, workflow steps, and parameter settings, thereby
facilitating transparency and reproducibility of workflow products. In
addition to this external use, provenance information can also be used
internally, e.g., to allow scientists to trace sources of errors and
to debug their workflows.

  The data provenance captured by workflow environments is sometimes
  called \emph{retrospective provenance} to distinguish it from
  another form called \emph{prospective provenance}
  \cite{Clifford2008original,Lim2010Prospective}. The former consists
  of data dependencies and lineage information recorded at runtime,
  which can then be used later for retrospective exploration and
  analysis (a.k.a.\ ``querying provenance''
  \cite{Davidson2008Provenance}). In constrast, \emph{prospective}
  provenance is a description of the computational process itself,
  i.e., the workflow specification is considered a form of provenance
  information, describing the \emph{method} by which analysis results
  and other data products are obtained. Scientific workflow systems
  therefore naturally support both forms of provenance, i.e.,
  prospective provenance by visually presenting a workflow as a
  directed graph with data and process steps, and retrospective
  provenance by capturing and subsequently exporting runtime
  provenance.

Despite these and other advanced features of workflow systems,
a vast number of computational ``workflows'' continue to be developed
using general purpose or specialized scripting languages such as
Python, \R, and \MATLAB. This is true in particular for the ``long
tail of science'' \cite{wallis2013if,Heidorn2008Shedding}, where
advanced features such as provenance support are rarely available.
For example, provenance libraries for \R\ have only recently been
announced \cite{Lerner2014RDataTracker}, while for Python, a new tool
called \NW\ has just been developed \cite{murta2014noWorkflow}. The
\NW\ (\emph{\textbf{n}ot \textbf{o}nly \textbf{workflow}}) system uses
Python runtime profiling functions to generate provenance traces that
reflect the processing history of the script. Thus, \NW\ allows users
to continue working in their familiar Python scripting environment,
without adopting a new system, while retaining the advantage of
automatic capture of retrospective provenance information similar to
the one available in workflow systems.

In the following, we describe a new tool called \YW\  that
complements \NW\ by revealing prospective provenance in
scripts, i.e., \YW\ makes latent workflow information from scripts explicit. In
particular dataflow dependencies that are often ``hidden'' inside of a
script and not easily understood by outsiders looking at the script
are extracted from simple user annotations and can then be exported
and visualized in graph form. 

\pagebreak

 \noindent The main features of \YW\ (or \yw\ for short) are:

\begin{itemize}

\item \yw\ exposes prospective provenance (workflow structure and
  dataflow dependencies) from scripts based on simple user annotations.

\item \yw\ annotations are embedded inside of comments, so they
  are \emph{language independent} and can be used, e.g., in Python, \R,
  and \MATLAB.

\item \yw\ annotations and the underlying model are deliberately kept
  simple to allow scientists a very low entry bar for adoption.

\item The \yw-toolkit is a grass-roots, agile, open source effort, whose
 simple and modular architecture and underlying UNIX philosophy
 facilitates interoperability and extensibility.

\item The current \yw\ prototype generates different, easily reusable
  output formats, including three different graph views, i.e., a
  \emph{process-centric}, a \emph{data-centric}, and a \emph{combined
    view} of the extracted workflow graph in Graphviz/DOT form.

 \end{itemize}

 \noindent We discuss \yw\ limitations and plans for future
 development in Section~\ref{sec-conclusions}.

\section{\YWT\ Model and Annotation Syntax}\label{sec-ywmodel}

In order to use the \YW\ tools, a script author marks up scripts using
a simple keyword-based annotation or tagging mechanism, embedded
within the comments of the host language. \yw\ annotations are
expressions of the form
\ywa{@\emph{tag}}~\textvisiblespace~\ywa{\emph{value}}. Here,
\ywa{@\emph{tag}} is one of the recognized \yw\ keywords, after which a
\ywa{\emph{value}} follows, separated by one or more whitespace
characters. Thus, the \yw\ annotation syntax mimics the syntax of
conventional documentation generators such as Javadoc and DOxygen.

The \yw\ tool then interprets the embedded, structured comments and
builds a simple workflow model of the script. This model represents
scripts in terms of scientific workflow entities, i.e., programs,
workflows, ports, and channels:

\begin{itemize}

\item A \emph{program block} (short: \emph{program} or \emph{block})
  represents a computational step in the script that receives input
  data and produces (intermediate or final) output data. A program is
  designated in a script by bracketing the relevant code between a
  pair of \ywa{@begin} and \ywa{@end} comments. Program blocks are
  usually visualized as boxes. A block that contains other programs is
  considered a \emph{workflow}. \item A \emph{port} represents a way
  in which data flows into or out of a program or workflow. Ports are
  identified by \ywa{@in} and \ywa{@out} annotations in the source
  code comments.

\item A \emph{channel} is a connection between an \ywa{@out} port
of a program and an \ywa{@in} port of another (or, in case of feedback
loops, the same) program. \yw\ infers
channels by matching the names of \ywa{@in} and \ywa{@out} 
ports within the same workflow.
\end{itemize}

 \begin{figure}[t]
   \centering
   \includegraphics[width=1.0\textwidth]{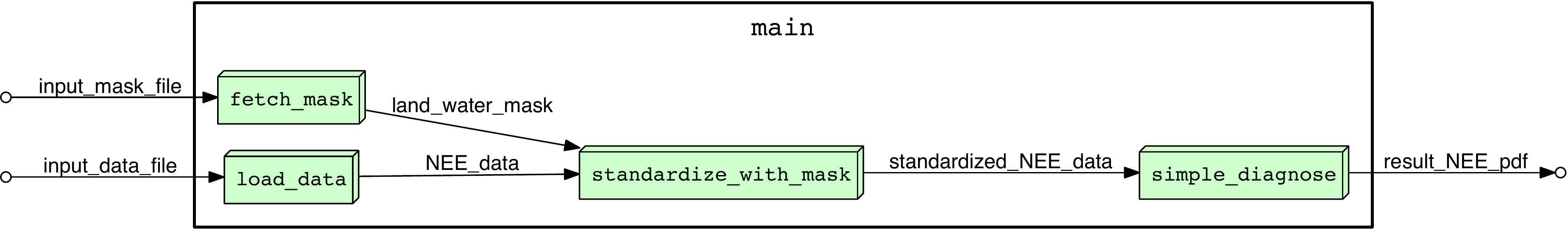}
   \caption{Process-oriented workflow view of a script: boxes
     represent \emph{programs} (code blocks); edges represent dataflow
     \emph{channels}; edge labels indicate \emph{data}
     elements.}
   \label{fig-simple-process}
 \end{figure}

 \noindent \figref{fig-simple-process} depicts a workflow view
 extracted from a sample Python script for standardizing Net Ecosystem
 Exchange (NEE) data in the MsTMIP project; cf.\
 Section~\ref{sec-MsTMIP}.

\begin{figure}[h]
   \centering
   \includegraphics[width=1.0\textwidth]{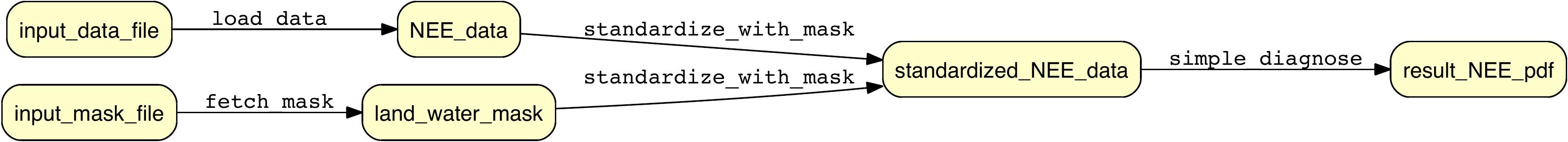}
   \caption{Data-oriented workflow view: program blocks are mentioned
     in edge labels only, while data channels are exposed as proper
     graph nodes.}
   \label{fig-simple-data}
 \end{figure}

\begin{figure}[h]
   \centering
   \includegraphics[width=1.0\textwidth]{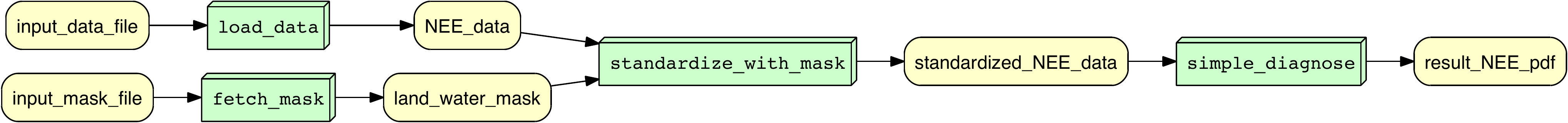}
   \caption{Combined workflow view of a script: both programs and data
   are nodes.}
   \label{fig-simple-combined}
 \end{figure}

\paragraph{Alternative Workflow Views.}
The process-oriented view in \figref{fig-simple-process} is the
default \yw\ view shown to the user, as it emphasizes the overall
block structure, given by the script author using \ywa{@begin} and
\ywa{@end} markers.  However, the extracted \yw\ model can also be
rendered in other forms.  For example, \figref{fig-simple-data}
depicts a \emph{data-oriented view}, where data elements (i.e.,
dataflow channels obtained from \ywa{@in} and \ywa{@out} tags) are
shown as nodes, while programs are only mentioned in edge labels.
Finally, \figref{fig-simple-combined} shows a \emph{combined workflow
  view}, i.e., in which both programs and data channels are
represented as nodes.

\section{Querying \YWT\ Models}\label{sec-querying}

The workflow structure of large scripts can be difficult to interpret
fully even when represented graphically.  While the \yw\
prototype is limited to such graphical views, the \yw\ comments and
model are sufficient to support queries that reveal specific aspects
of the script in workflow terms.  Example
workflow-structure queries that will be supported by \YW\ include:
\begin{itemize}
\item List all of the code blocks defined in the script along with any description given for each.
\item List the code blocks nested (directly or indirectly) within a particular code block.
\item List the code blocks that invoke a particular function or external program.
\item List the code blocks that contain a particular block (directly or indirectly).
\item List the code blocks that receive inputs derived (directly or indirectly) from the outputs of a particular upstream code block.
\item List the code blocks affected (directly or indirectly) by a particular parameter value provided to the script.
\end{itemize}

\paragraph{Prospective Data Provenance Queries.}
\YW\ additionally will allow scripts
marked up with \yw\ comments to be queried from a data provenance
perspective. Because \YW\ analyzes the definition of a workflow (the
script plus \yw\ comments) rather than information recorded during a
run of the script,  \YW\ will support queries
against \emph{prospective} provenance.
Example prospective provenance queries include:
\begin{itemize}
\item Given the name of an output of the script, list the inputs to
  the script that the output depends on (directly or indirectly). 
\item List the computational steps (code blocks) involved in deriving
  a particular output of the script, or of a named intermediate data
  product. 
\item For a particular computational step reveal where each input to
  the step comes from: an input to the script, a constant in the
  script, a value produced by a different step, etc. 
\item Reveal the complete derivation of a particular script output.
  That is, list the sequence of code blocks and input and intermediate
  data products leading to the output. Results of queries of this kind
  optionally may be rendered graphically. 
\end{itemize}

\paragraph{Inference of Retrospective Data Provenance.}
As described above, \YW\ will allow prospective provenance to be
inferred from scripts marked up with \yw\ comments.  We additionally
foresee that combining the information extracted from a marked-up
script with references to data files corresponding to a run of that
script will in some cases allow the retrospective provenance of those files to be
inferred (see also \cite{Bowers2012Provenance} and
\cite{Zinn2010Abstract}). That is, in cases where the entire sequence of data
derivation steps for a particular output can be determined unambiguously from YW
annotations, \YW\ will support queries of the following kind even in the absence of a
run-time data-provenance recorder:

\begin{itemize} 

\item Given a file output by a run of a script, indicate which files
  input to the script this output file was derived  from (or affected
  by). 

\item Given an input file to a script, indicate which  output files
  were derived (or affected) by the data contained in that file.

\item Indicate which parameter values applied to a run of the
  script affected which of its output files. 

\end{itemize}

\section{\YWT\ Examples}

 \begin{figure}[t]
   \centering
   \includegraphics[width=.99\textwidth]{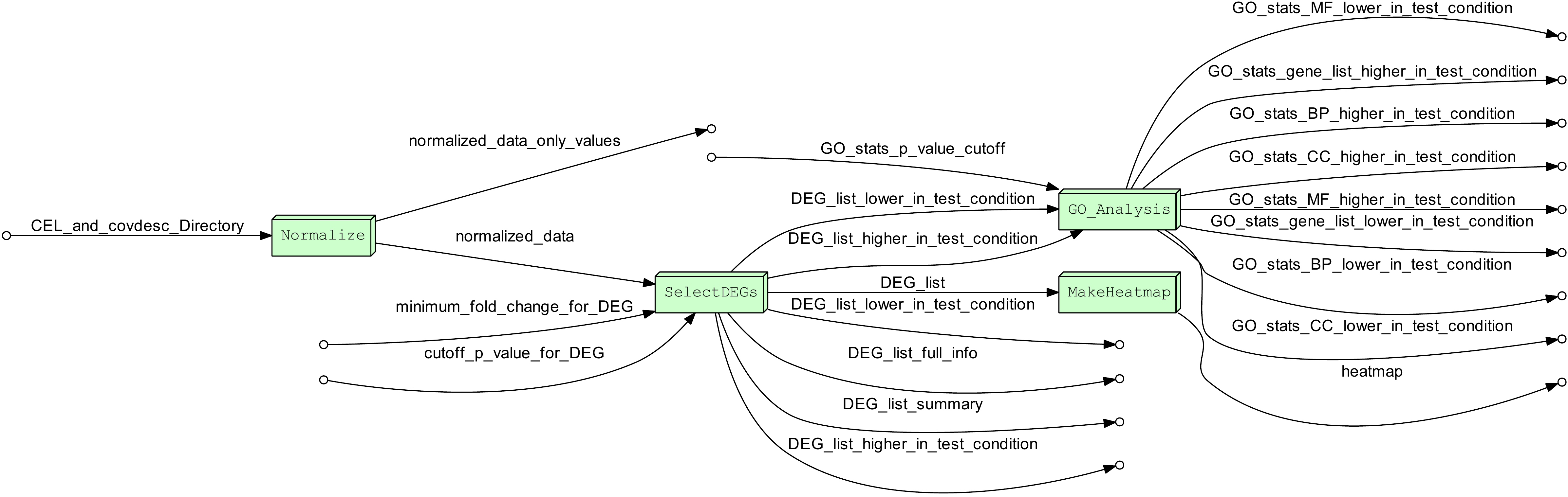}
   \caption{Process workflow view of an
     Affymetrix analysis script (in \R).}
   \label{fig-affymetrics}
 \end{figure}

 In the following we show \YW\ views extracted from real-world
 scientific use cases.  The scripts were annoted with \yw\ tags by
 scientists and script authors, using a very modest training and
 mark-up effort.\footnote{For all of these scripts, learning the \yw\
   model and annotating the scripts was done in a few hours.}  Due to
 lack of space, the actual \MATLAB\ and \R\ scripts with their \yw\
 markup are not included here. However, they are all available from
 the  \href{http://yesworkflow.org/yw-idcc-15}{yw-idcc-15} repository 
on the \yw\ GitHub site \cite{YWgithub2015}.

\subsection{Analysis of Gene Expression Microarray Data}

 Bioinformatics workflows commonly possess a pattern of large numbers
 of incoming parameters and outputs at each stage of computation. In
 addition, analysis of even a single bioinformatics dataset tends to
 yield a large number of different output files. Hence, bioinformatics
 pipelines are attractive candidates for workflow systems, which can
 capture this complexity \cite{bieda2012kepler}. \figref{fig-affymetrics}
 shows a \YW\ representation of an \R\ script performing a
 classic, complex bioinformatics task: analysis of Affymetrix gene
 expression microarray data. This \R\ script was modeled on our previous
 workflows developed in the Kepler environment \cite{stropp2012workflows}.
 The script analyzes experiment designs consisting of two conditions
 (e.g., microarrays from control-treated cells vs microarrays from
 drug-treated cells) with multiple replicates in each condition. The
 \R\ script employs a set of standard BioConductor
 \cite{gentleman2004bioconductor} packages mixed with custom programming. The
 workflow consists of four fundamental tasks: normalization of data
 across microarray datasets (\ywm{Normalize}), selection of
 differentially expressed genes (\ywm{DEGs}) between conditions
 (\ywm{SelectDEGs}), determination of gene ontology (\ywm{GO})
 statistics for the resulting datasets (\ywm{GO\_Analysis}), and
 creation of a heatmap of the differentially expressed genes
 (\ywm{MakeHeatmap}). Each module produces outputs, and each module
 (aside from \ywm{MakeHeatmap}) requires external parameter
 inputs. Importantly, this graphical representation clearly indicates
 the dependence of each module on datasets and parameter inputs. This
 example demonstrates that \YW\ can provide informative visualizations
 of bioinformatics workflows, especially workflows involving large
 numbers of inputs and outputs.

\subsection{Terrestrial Biospheric Modeling}\label{sec-MsTMIP}

  \begin{figure}[t]
   \centering
   \includegraphics[width=.99\textwidth]{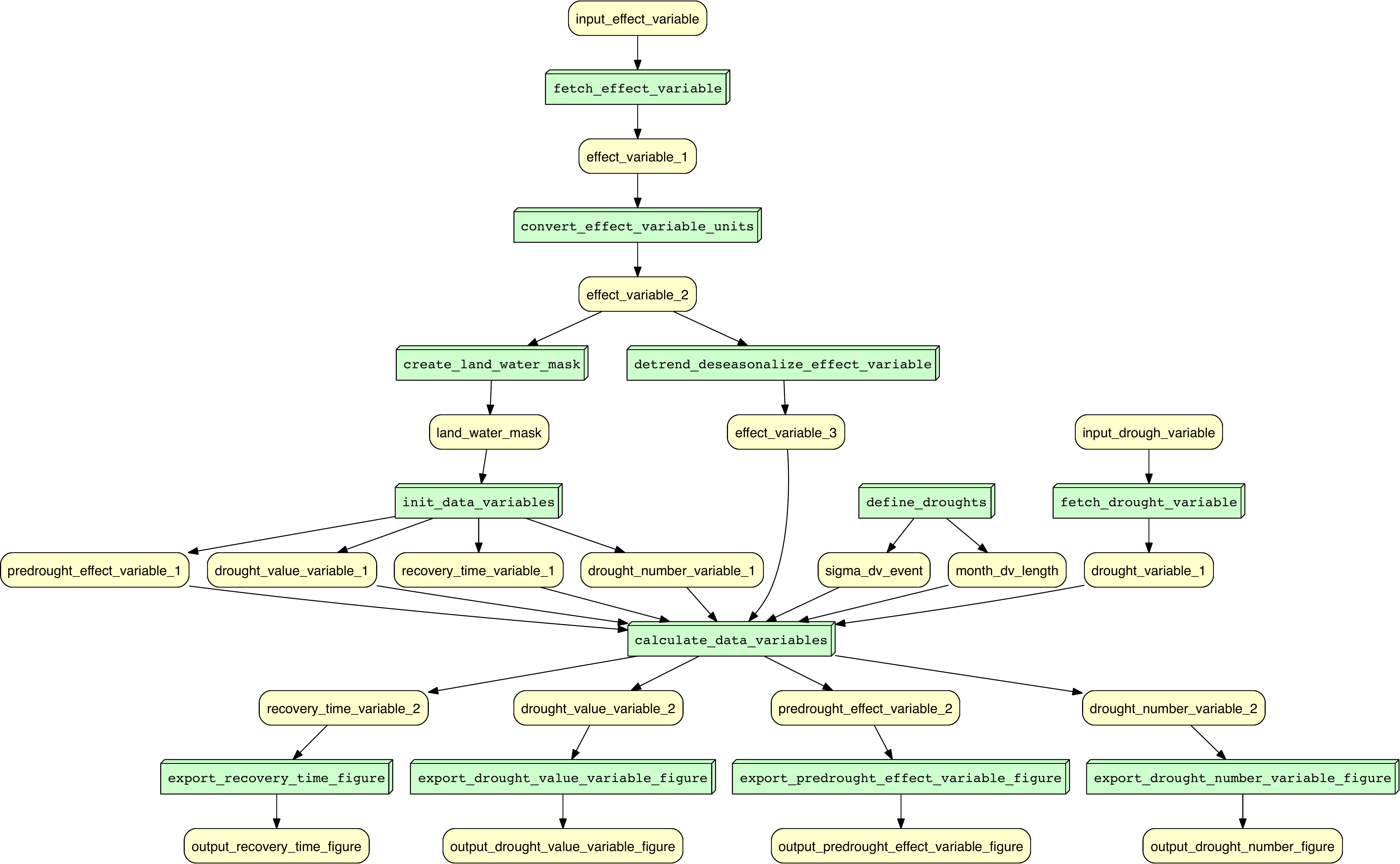}
   \caption{Combined workflow view of a MsTMIP script (in
     \MATLAB). \yw\ views can be easily tweaked via Graphviz
     properties in the generated DOT files: here, a ``Taverna-style''
     top-down layout is used, as opposed to the default left-to-right
     display.}
   \label{fig-mstmip}
 \end{figure}

In the Multi-scale Synthesis and Terrestrial Model Intercomparison
Project (MsTMIP)\footnote{\url{http://nacp.ornl.gov/MsTMIP.shtml}},
climate scientists primarily use \MATLAB\ scripts to standardize
terrestrial biosphere model output across multiple models and
simulation runs for intercomparison purposes and to facilitate
diagnosis and attribution.
MsTMIP is a large, collaborative effort, aimed at harmonizing a number
of complex terrestrial biospheric models for the purposes of comparing
these model outputs \cite{huntzinger2013north}. There is a strong need
to standardize many aspects of the MsTMIP process, to assure greater
uniformity in the treatment of the codes and outputs of the disparate
models in the intercomparison analyses. Current practice in MsTMIP,
however, is representative of many scientific investigations, i.e.,
researchers develop their codes with a specific focus on functionality
and efficiency. Comments are added primarily as ``bookmarks'' to
assist with accessing appropriate code areas for debugging,
optimization, or discussion. In the more general case, depending on
whether the codes are developed in a collaborative context, structured
in-code documentation may be recommended or required by the
project. Nevertheless,  the mechanisms for these ``code annotations'' are
typically unformalized and unstructured, and rely primarily on the
ability to insert non-executable ``comment'' statements in the code.

As the complexity of code grows, and the numbers of variants and
alternative approaches increases, MsTMIP researchers need a clear and
consistent way to document, review, and share their model
intercomparison scripts. This provides a compelling use case for \yw,
in that MsTMIP brings together models from a number of independent
efforts that require harmonization into a single framework for
evaluating their relative capabilities to predict critical earth
system features, such as global Net Ecosystem Exchange (NEE) data from
terrestrial biogeographic realms.

 \subsection{Paleoclimate Reconstruction}

 \begin{figure}[t]
   \centering
   \includegraphics[width=.7\textwidth]{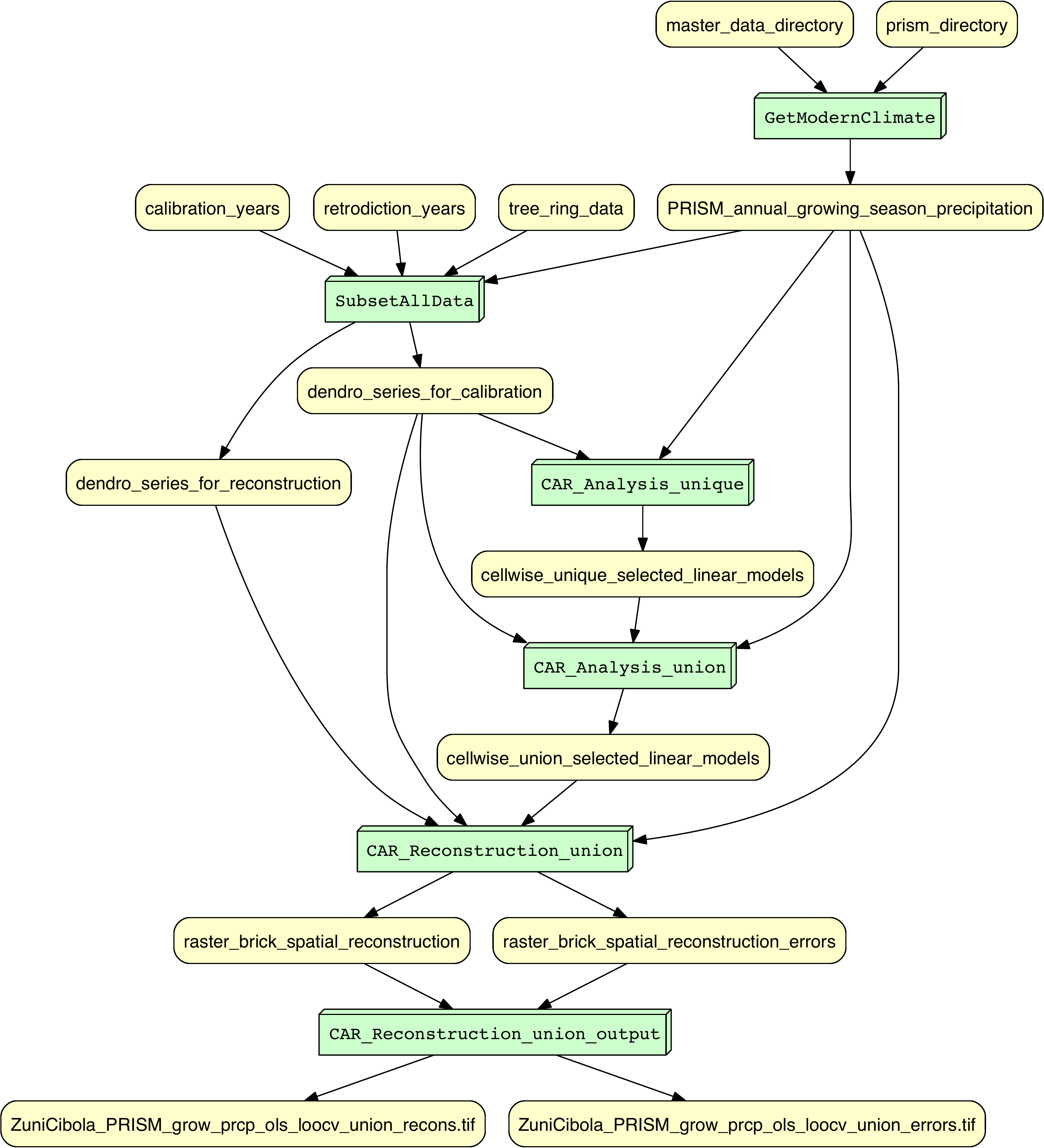}
   \caption{Combined workflow view of a paleoclimate reconstruction
     \R\ script \protect\cite{bocinsky2014}.}
   \label{fig-skope}
 \end{figure}

 As another working example from a different field, we have used the
 \YW\ markup syntax to analyze the paleoclimate reconstruction
 workflow presented by Bocinsky and Kohler \cite{bocinsky2014}. Their
 reconstruction method takes as input a spatial interpolation of
 contemporary weather data, the long-term record of climate held in
 regional tree-ring chronologies, and a handful of parameters, and
 uses a novel regression-based analysis method to generate spatial
 reconstructions of climate extending 2000 years or more back in
 time. \figref{fig-skope} shows that the \yw\ system nicely exposes
 the prospective provenance hidden in the underlying \R\ script, even
 for scripts whose workflow views are highly non-linear.

\section{\ywt\ Architecture}

The \YW\ software distribution is envisioned as a set of standard
modules that can be used together or independently. The primary goal
of this modularity is to enable \yw\ users and developers
independently to implement alternatives to any module, as needed, to
solve problems particular to their research domain. It will be
possible to develop these alternative implementations and extensions
in any programming language.  One way we plan to facilitate such easy
replacement of \yw\ modules is to require that each standard module
optionally input and output files--with well-defined
formats--representing the expected inputs or outputs of that
module. Any program that produces or consumes these file formats can
then function as an alternative to one or more standard \yw\ modules
and can provide identical, overlapping, or completely different
capabilities (e.g., the current \textsf{YW} prototype is primarily
implemented in Java, but also contains some alternative \textsf{YW}
modules implemented in Python).

Five standard modules (implemented in Java) currently are implemented
or planned: The \ywm{YW-Extract} module identifies \yw\ comments in a
script and produces a language-independent representation of the
script and the YW annotations. \ywm{YW-Model} interprets the comments
identified by \ywm{YW-Extract} and builds a model of the script in
terms of entities analogous to the components of a traditional
scientific workflow as described in Section~\ref{sec-ywmodel}, while
\ywm{YW-Graph} operates on the outputs of \ywm{YW-Model} to produce
the dataflow graphs discussed in that same section. As described in
Section~\ref{sec-querying}, the planned \ywm{YW-Query} module will
allow users to probe the structure of a complex script without having
to inspect a visual representation of it. An envisioned
\ywm{YW-Validate} module will ensure that \yw\ comments in a script
are consistent both with the other \yw\ comments in the script and
with the script itself. Finally, the \ywm{YW-CLI} module enables a
user to execute sequences of the standard modules, starting from an
input file with format appropriate to the first module in the executed
sequence.

\section{Related Work}

The \yw\ approach can be seen in the tradition of programming code
annotation, which is widely used for facilitating code understanding
and for generating documentation (e.g.,
DOxygen\footnote{\href{http://www.doxygen.org}{www.doxygen.org}},
Epydoc\footnote{\href{http://epydoc.sourceforge.net}{epydoc.sourceforge.net}},
Javadoc\footnote{\href{http://www.oracle.com/technetwork/java/javase/documentation/index-jsp-135444.html}{www.oracle.com/technetwork/java/javase/documentation/index-jsp-135444.html}
}, etc.) \yw\ builds on programming code annotation to provide a
higher level of abstraction by revealing the dataflow that underlies the
interactions between the different pieces of a script or program.

\yw\ is also related to ideas from literate programming\footnote{Don
  Knuth has argued \cite{knuth1984literate} that we should change our
  traditional attitude to programming: ``Instead of imagining that our
  main task is to instruct a \emph{computer} what to do, let us
  concentrate rather on explaining \emph{human beings} what we want a
  computer to do''.} and available in tools such as
Knitr~\cite{xie2013dynamic} and IPython~\cite{perez2006ipython}. In
literate programming, a script is decomposed into snippets of macros,
which are interspersed within documents that are written in natural
language to explain the scripts and eventually analyze the results it
generates upon execution. While borrowing ideas from literate
programming, \yw\ is primarily targeted for developers who are using
pure traditional scripting environments to edit their scripts and
programs. \yw\ aims at providing a consistent interpretation and
visualization of codes wherever the language provides for insertion of
non-executable ``comments''.

\yw\ can also contribute to the area of reproducible computational
research \cite{stodden2014implementing}, which seeks to provide
scientists with sufficient information to understand and eventually
validate the results claimed by their peers. For instance, the SOLE
system \cite{pham2012sole} allows linking articles with science
objects, which can be source code, a dataset, or a workflow. SOLE
allows the reader (curator) to specify human-readable tags that link
the paper with science objects, and it transforms each tag into a URI
that points to a representation of the corresponding object. While in
SOLE the scientific article is the main object that contains links to
other (science) objects, we focus on the scripts produced by the
scientists, and aim to facilitate the understanding of their dataflow
logic.  Gavish and Donoho~\cite{gavish2011Universal} present the
notion of a \emph{Verifiable Computational Result} (VCR), where every
result is assigned a unique identifier, and results produced under the
exact same conditions have the same identifier to support
reproducibility.

Various tools have been proposed to capture the runtime provenance of
scripts. Mechanisms that capture provenance at the operating system
level~\cite{frew2008Automatic,guo2012BURRITO,muniswamy2006Provenance}
monitor system calls to track the data dependencies between
computational processes. Some
tools~\cite{bochner2008Python,davison2012Automated,huq2013ProvenanceCurious,murta2014noWorkflow}
have been developed to capture runtime provenance for Python scripts:
while Bochner et al.~\cite{bochner2008Python} and
Davison~\cite{davison2012Automated} propose Python libraries and APIs that
need to be added to the code to capture the execution steps,
ProvenanceCurious~\cite{huq2013ProvenanceCurious} and
noWorkflow~\cite{murta2014noWorkflow} are transparent and do not
require changes to the scripts. Similarly,
RDataTracker~\cite{Lerner2014RDataTracker} captures provenance from
the execution of \R\ scripts, and the approach taken by Tariq et
al.~\cite{Tariq2012Towards} supports all programming languages allowed
by the LLVM compiler framework. We note that the \yw\ approach is
complementary to these tools, since it captures prospective provenance
of scripts. We argue that \yw, along with runtime provenance
approaches, provide a low-effort entry point for scientists who want
to reap some of the benefits of scientific workflow systems while
still using their familiar scripting environments.

\section{\YWT\ Development Roadmap}\label{sec-conclusions}

In the following we list some limitations of the current \yw\
prototype and highlight features planned for future
releases of the software.

\paragraph{Visualization of Nested Code Blocks.}
The \ywm{YW-Extract} and \ywm{YW-Model} modules support
 nesting of code blocks. Any pair of \ywa{@begin} and \ywa{@end}
comment lines can enclose code that contains any number of other code
blocks delimited with \ywa{@begin} and \ywa{@end} comment lines.
The workflow model constructed for a script reflects such nesting, i.e.
the top-level workflow corresponding to the script as a whole may contain 
one or more programs (code blocks), and any of these programs can in turn be a 
sub-workflow that contains further nested programs and workflows.
Future versions of \ywm{YW-Graph} will reveal these nested code blocks 
and render sub-workflows graphically.

\paragraph{Functions and Function Calls.}
\ywm{YW-Extract} currently expects nested code blocks to be defined
in-line. However, many scripts are structured as functions (or
classes) with a top-level script that calls these functions (or
methods on objects). These functions can in turn call other functions.
Future versions of \YW\ will allow function declarations to be marked
up with \yw\ comments in a manner similar to that supported by Javadoc and
DOxygen. Calls to these functions also will be annotated with \yw\
markup. The result will be that \ywm{YW-Extract} and \ywm{YW-Model}
will be able to represent function calls as nested code blocks.

\paragraph{Interactive Graphs.}
\ywm{YW-Graph} currently produces static graphical views (in the
well-known Graphviz-DOT format). 
An interactive viewer for \yw\ graphical
output will make these graphs easier to explore and interpret. In the planned
graphical user interface, clicking on a data item in the combined or
data views optionally will highlight the (prospective) direct and
indirect data dependencies for that data item (the data from which it
will be derived when the script is run).  Features for expanding and
collapsing nested subworkflows also will facilitate exploration of
these graphs.

\paragraph{Live Graph View.}
Although the primary function of \YW\ is to reveal workflow-like
structure in existing scripts, \YW\ also can be used as a \emph{design
  tool} when developing new scripts (or even before a script is
written). Future versions of \YW\ will better support such
applications by providing live-update features to the interactive
graph capabilities described above. Given a set of script files, the
live-graph feature will monitor these files for changes and update the
chosen graphical view automatically. Users of this feature will
continue to be able use their favorite text editor or IDE for
developing their scripts.

\paragraph{Distinguished Data and Parameters.}
The inputs to scripts for processing scientific data often can be
viewed either as data (the data to be processed by the scripts) or as
parameters (values that control how that data is processed).
Planned versions of the \yw\ comment vocabulary will allow data and parameters
to be distinguished. \ywm{YW-Graph} optionally will emphasize graph edges,
nodes, and labels representing data over those representing
parameters.

\paragraph{Validation of Comments.}
The future \ywm{YW-Validate} module will perform
extensive validation of \yw\ comments in light of the actual code in
the script.  This capability will help guide users adding \yw\
comments to their script. Perhaps more importantly, automatic
validation will help prevent initially correct \yw\ comments from
becoming stale (i.e., incorrect) when the underlying script is changed
or refactored.
Validity checks that \ywm{YW-Validate} will perform include:
\begin{itemize} 
\item Confirm that data names used in \ywa{@in} and \ywa{@out} comments
actually appear in the code bracketed by associated \ywa{@begin} and  \ywa{@end}
  comments.
\item Confirm that the names of functions referred to in \yw\ comments
  for function declaration or for function calls match the names of
  the functions actually declared or called.
\item Confirm that  continuous data dependency chains exist from each script output all
  the way back to script inputs (and embedded constants).
\end{itemize}

\section{Conclusions}

\YW\ is an agile, grass-roots effort that aims at bringing workflow
modeling and analysis features to scientific ``workflows'' that are
defined in script form. Through simple user-annotations in the
comments of scripts, dataflow and workflow structure are revealed by
the \yw\ toolkit. The user can thus exploit prospective 
provenance information from scripts, e.g., by
visualizing, querying, and analyzing this information.

Our early \yw\ prototype \cite{YWgithub2015} has been used by
scientists from different domains to mark up complex, real-world
scientific scripts with ease. Encouraged by the enthusiastic response
of the early adopters, a number of researchers will be incorporating
\yw\ into their projects, thereby guiding and driving the future
development of \YW.

MsTMIP researchers plan to annotate their scripts such that authors,
as well as reviewers and potential new users, will be able to click on
the workflow steps in the interactive \yw\ graph viewer and inspect
the corresponding code-blocks in the original script. When clicking on
data elements, they will be taken to a folder containing the data
instances that were used in the various runs of the script (provided
these have been shared). Since the \yw\ approach is language
independent, it will also facilitate code migration, say from \MATLAB\
to \R, or from \R\ to Python.

In the Kurator project \cite{kuratorproject} we plan to enable
collection managers to author their own data curation workflows using
both an Akka-based workflow system and via scripting languages such as
Python and \R. In the latter case, Kurator tool users will annotate
their scripts with \yw\ comments to enable provenance queries to span
script-based curation workflows.
The Kurator team also plans to use the \ywm{YW-Graph} and
\ywm{YW-Query} tools to graphically render workflows defined using the
Kurator-Akka workflow system and to query the prospective provenance
of products of these workflows.

Finally, DataONE is planning a number of enhancements to the \yw\
annotation language. For example, in addition to the currently
supported, simple user-defined vocabulary for program blocks and data
elements, controlled vocabularies from shared ontologies may be used
with these extensions. Similarly, to improve \yw\ interoperability
within the DataONE infrastructure, PROV \cite{moreau2013prov} and
ProvONE \cite{dataone2014provone} compatible vocabulary extensions
may be used in \YW\ in the future.

\paragraph{Acknowledgements.}
Work supported in part by the National Science Foundation under awards
DBI-1356751 (Kurator), ACI-0830944 (DataONE), SMA-1439603 (SKOPE).

\small

\bibliographystyle{alpha-initials-big}
\bibliography{yw-idcc15-arxiv}
\end{document}